\documentstyle[preprint,aps,psfig]{revtex}
\def\be{\begin{equation}}
\def\ee{\end{equation}}
\def\bea{\begin{eqnarray}}
\def\eea{\end{eqnarray}}
\begin{document}
%\begin{frontmatter}
\title{Distinguishing Hadronic Cascades from Hydrodynamic Models in
Pb(160~AGeV)+Pb Reactions by Impact Parameter Variation}

\author{M.~Bleicher${}^{a,e}$, M. Reiter${}^a$,
A. Dumitru${}^{b,g}$,
J. Brachmann${}^a$,
C. Spieles${}^{c,f}$,
S.A. Bass${}^{d,f}$,
H.~St\"ocker${}^a$, 
W.~Greiner${}^a$}

\address{${}^a$ Institut f\"ur
Theoretische Physik,  J.~W.~Goethe-Universit\"at,\\
60054 Frankfurt am Main, Germany}

\address{${}^b$ Department of Physics, Yale University,\\
        New Haven, Connecticut, USA}

\address{${}^c$ Nuclear Science Division,\\
        Lawrence Berkeley National Laboratory,\\
        Berkeley, CA 94720, USA}

\address{${}^d$ Department of Physics, Duke University,\\
        Durham, N.C. 27708-0305, USA}
\footnotetext{E-mail: bleicher@th.physik.uni-frankfurt.de}
\footnotetext{${}^e$ Fellow of the Josef Buchmann Foundation}
\footnotetext{${}^f$ Feodor Lynen Fellow of the Alexander v. Humboldt Foundation}
\footnotetext{${}^g$ Supported by a postdoctoral fellowship of the
Deutscher Akademischer Austauschdienst (DAAD)}


\maketitle
\begin{abstract}
We propose to study the impact parameter dependence of the
$\overline\Lambda /\overline p$ ratio in Pb(160AGeV)+Pb reactions.
The $\overline\Lambda /\overline p$ ratio is a sensible tool to
distinguish between hadronic cascade models and hydrodynamical models,
which incorporate a QGP phase transition. 
\end{abstract}

%\end{frontmatter}
\newpage

Hadron abundances and ratios have been suggested as possible signatures
for exotic states and phase transitions in hot and dense nuclear matter.
Bulk properties like temperatures, entropies and chemical potentials
of highly excited hadronic matter have been extracted from high energy
heavy ion data assuming thermal and chemical equilibrium
\cite{stoecker,braun-munzinger,cleymans,rafelski}.
However, unambiguous signals of a phase transition into an equilibrated 
deconfined quark gluon plasma (QGP) state are still missing: 
the predicted change in observables, e.g. 
strangeness enhancement \cite{rafelski}, may also be understood in hadronic
non-equilibrium transport models \cite{rqmd-raffi}.

In this letter, the variation of the $\overline\Lambda /\overline p$
ratio\footnote{Note that the $\overline\Lambda$'s contain the 
decayed $\overline{\Sigma^0}$'s, while $\overline p$ do not contain
decays from $\overline\Lambda$($\overline{\Sigma^0}$)'s.} as a function of the impact parameter b
in Pb(160~AGeV)+Pb reactions is proposed as a method to distinguish
equilibrium from non-equilibrium scenarios.
In models based on the assumption of
local thermal equilibrium (e.g. in thermal models or hydrodynamical
models) the $\overline \Lambda / \overline p$
ratio is sensitive only to the temperature and chemical potentials 
achieved in the reaction.  Microscopic transport
theory, however, is not constrained by equilibrium assumptions. 
Thus, this ratio provides a sensible tool to probe the creation of a
chemically equilibrated phase in nucleus-nucleus collisions as a function of
centrality. 
A comparison to upcoming data by the NA49 and the CERES collaborations
may therefore provide an estimate of the degree of local
chemical equilibration. The differences in the predicted centrality 
dependence among the discussed models can
help to determine the applicability of these theories. 

In the 3-fluid hydrodynamical model \cite{3f} \cite{Manuel-S/A} 
an equation of state with a first order phase transition to a QGP is used. 
We employ that model to calculate entropy production during the initial
stage of the reaction as described in detail in~\cite{Manuel-S/A}.
To show the behaviour of the $\overline\Lambda /\overline p$  ratio 
for the chemical equilibrium case, the creation of a fireball, composed of all
hadrons up to mass $m=2$~GeV, with a uniform
$S/A$ ratio (entropy $S$ per net
participating baryon $A$) 
and net baryon density $\rho$ is assumed. 
The 3-fluid model is used to calculate $S/A$ 
as a function of impact parameter b.
The hadron ratios are calculated assuming
chemical isochronous freeze-out at a net baryon density $\rho=\rho_0/2$.

\begin{figure}[t]
\vskip 0mm
\vspace{-1.0cm}
\centerline{\psfig{figure=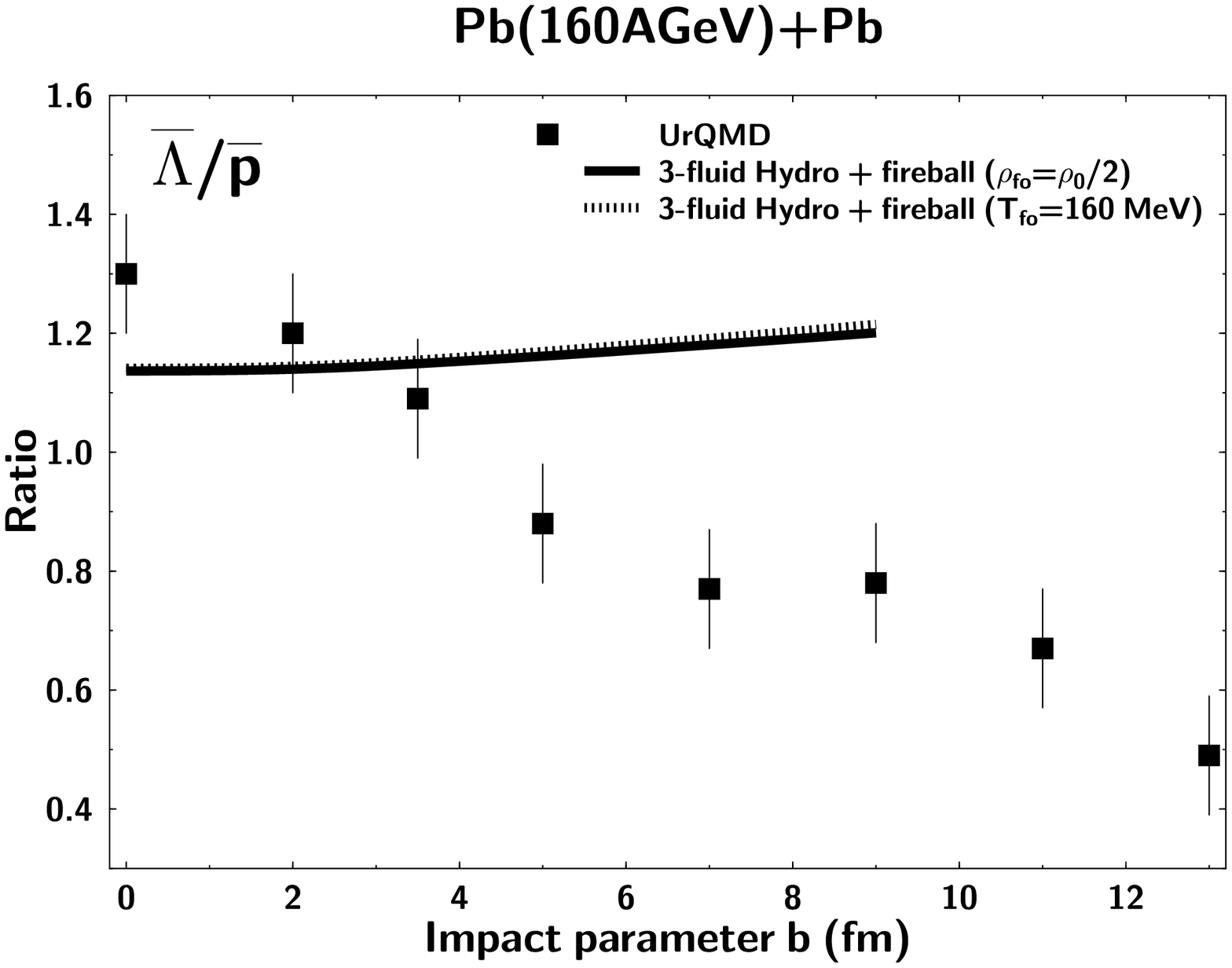,width=20cm}}
\vskip 2mm
%\vspace{-1.0cm}
\caption{Impact parameter dependence of the $\overline\Lambda
/\overline p$ ratio in Pb(160~AGeV)+Pb reactions. The full squares denote
hadronic cascade (UrQMD) calculations, the lines show the 3-Fluid Hydro + Fireball 
calculation including a first order phase transition (full
line: freeze-out at $\rho=\rho_0/2$, dashed line: freeze-out at $T=160MeV$).
\label{alap}}
\end{figure}

For $S/A>25$, which is the relevant $S/A$ range for SPS energies, 
the $\overline\Lambda /\overline p$ ratio is practically unchanged if 
a chemical freeze-out temperature of $T=160$~MeV 
(as suggested in~\cite{braun-munzinger}) is 
employed instead of the
freeze-out density $\rho=\rho_0/2$ (cf. Fig.~\ref{alap}).

Within the UrQMD model \cite{urqmd}, the
non-equilibrium dynamics is treated in a microscopic hadronic scenario.
Baryon-baryon, meson-baryon and meson-meson collisions 
lead to the formation and decay of resonances and color flux tubes. The
produced, as well as the incoming particles, 
rescatter in the further evolution of the system.

Fig. \ref{alap} shows the $\overline\Lambda /\overline p$ ratio for
different impact parameters from 0~fm to 13~fm (UrQMD) and 0~fm to 9~fm
(3-fluid Hydro+Fireball; for larger impact parameters, the 
assumption of a chemically equilibrated fireball may not be justified).
The hydrodynamical calculation with phase transition
is depicted by the black line (using a freeze-out 
density $\rho=\rho_0/2$) and the
dotted line (using a freeze-out temperature $T=160$~MeV), 
the full squares denote the - microscopic 
non-equilibrium - UrQMD \cite{urqmd} calculation.

In the 3-fluid  approach the $\overline\Lambda /\overline p$  
ratio stays constant with b.
 
In contrast, the hadronic UrQMD model yields a strong
dependence of this ratio on impact parameter b. 
The $\overline\Lambda/\overline p$ ratio drops rapidly with 
increasing b from 1.3 to 0.5. 

The behaviour of the 3-fluid hydrodynamical model 
can be understood in terms of
the local equilibration of the hot and dense medium. The 
temperature and specific entropy of the 
fluid elements
are only slightly affected by b
(at least for not too large impact parameters). 
As long as the assumption of local 
chemical equilibrium is justified, the
particle number densities are independent of 
the volume (in the grand canonical formulation) -
thus their ratios remain constant.

In the case of the microscopic UrQMD model, there is an interplay
between particle production and subsequent annihilation:
In peripheral (large b) collisions the $\overline\Lambda$ production is
basically the same as in proton+proton reactions. $\overline\Lambda$'s
and $\overline p$'s are produced via the fragmentation of  color flux
tubes (strings). The production of (anti-)strange quarks in the color
field is suppressed due to the mass difference between strange and up/down
quarks. This results in a suppression of $\overline\Lambda$ over
$\overline p$ by a factor of 2 ($\overline\Lambda /\overline p \approx
0.3-0.5$ in pp).

In central Pb+Pb encounters meson-baryon and meson-meson reactions work
as additional sources for the anti-hyperon and anti-proton production. The
absolute $\overline\Lambda$, $\overline p$ yield increases far above the
naive p+p extrapolation. Then, additional rescattering effects have
to be taken into account in the hot and dense medium. Anti-baryons are
strongly affected by the comoving baryon density.
The additive quark model yields smaller 
annihilation cross section of anti-lambdas 
than the annihilation cross sections of anti-protons at 
the same momentum: Thus the annihilation probability for $\overline\Lambda$'s  
is smaller than for $\overline p$'s, leading to an increase of the
$\overline\Lambda /\overline p$ ratio above 1 in very central reactions.

In conclusion, it has been demonstrated that a study of the impact parameter
dependence of the $\overline\Lambda /\overline p$ ratio can be used as 
a powerful tool to distingush local equilibrium from off-equilibrium 
models for heavy ion collisions.

\section*{Acknowledgements}
This work is supported by the BMBF, GSI, DFG and Graduiertenkolleg
'Theoretische und experimentelle Schwerionenphysik'.

\newpage

\end{document}